\documentclass[aps,prc,twocolumn,superscriptaddress]{revtex4}

\usepackage{amssymb}
\usepackage{amsmath}
\usepackage{lscape}

\usepackage{varwidth}
\usepackage{multirow}
\usepackage{dcolumn}
\usepackage{tabularx}
\usepackage{booktabs}
\newcolumntype{C}{>{\centering\arraybackslash}X}

\usepackage{graphicx}
\usepackage{epstopdf}
\usepackage{graphics}
\usepackage{graphicx}
\usepackage{epsfig}
\usepackage{epstopdf}

\graphicspath{{figures/}}

\newcommand{\dd}{\ensuremath{{\rm d}}}
\newcommand{\pT}{\ensuremath{p_{\rm T}}}
\newcommand{\Ncoll}{\ensuremath{N_{\rm coll}}}
\newcommand{\Npart}{\ensuremath{N_{\rm part}}}
\newcommand{\TAA}{\ensuremath{T_{\rm AA}}}
\newcommand{\RAA}{\ensuremath{R_{\rm AA}}}
\newcommand{\sNN}{\ensuremath{\sqrt{s_{\rm NN}}}}
\newcommand{\s}{\ensuremath{\sqrt{s}}}

\newcommand{\dndy}{\ensuremath{\dd N/\dd y}}

\newcommand{\Zn}{\ensuremath{{\rm Z}^{0}}}
\newcommand{\Wpm}{\ensuremath{{\rm W}^{\pm}}}
\newcommand{\Wp}{\ensuremath{{\rm W}^{+}}}
\newcommand{\Wm}{\ensuremath{{\rm W}^{-}}}
\newcommand{\WZ}{\ensuremath{\Wpm / \Zn}}

\newcommand{\PYTHIA}{\textsc{Pythia}}

\newcommand{\cent}[2] {$#1$--$#2\%$}
\newcommand{\avg}[1]{\left\langle #1 \right\rangle}
\newcommand{\abs}[1]{\ensuremath{\left|#1\right|}}

\usepackage[dvipsnames]{xcolor}

\begin{document}

\title{Centrality dependence and isospin effect on $\Wpm$ and $\Zn$ productions \\
  in nucleus-nucleus collisions at $\sNN=5.02$~TeV}
\author{Dai-Mei Zhou$^1$ \footnote{zhoudm@mail.ccnu.edu.cn},
Yu-Liang Yan$^2$ \footnote{yanyl@ciae.ac.cn}, Liang Zheng$^3$,
Ming-Rui Zhao$^2$, Xiao-Mei Li$^2$, Xiao-Ming Zhang$^1$, Gang Chen$^3$,
An-Ke Lei$^1$, Ya-Qian Zhu$^1$, Xu Cai$^1$, and Ben-Hao Sa$^{1,2}$ \footnote{sabh@ciae.ac.cn}}
\affiliation{$^1$ Key Laboratory of Quark and Lepton Physics (MOE) and
Institute of Particle Physics, Central China Normal University, Wuhan 430079, China.\\
  $^2$ China Institute of Atomic Energy, P. O. Box 275 (10), Beijing, 102413 China. \\
  $^3$ School of Mathematics and Physics, China University of Geosciences (Wuhan), Wuhan 430074, China.}

\begin{abstract}
In this paper, the centrality dependent $\Zn$ and $\Wpm$ production and the
isospin effect in $\Wpm$ production are investigated with a parton and hadron
cascade model PACIAE in Pb--Pb collisions at $\sNN=5.02$~TeV. ALICE data of
$\Zn$ production in Pb--Pb collisions at $\sNN=5.02$~TeV are found to be
reproduced fairly well. The prediction on $\Wpm$ production in the same
collision system is given as well.
An interesting isospin effect is observed in exploring the charge asymmetry
between $\Wp$ and $\Wm$ as a function of the asymmetry between number of
valence $u$- and $d$-quarks varied from small to large collision
systems at center-of-mass energy 5.02 TeV. The results serve as a important
benchmarks for understanding the initial conditions of heavy-ion collisions.
\end{abstract}

\maketitle

\section {Introduction}
$\Wpm$ and $\Zn$ vector bosons are heavy particles with masses of $m_{\Wpm}=80.39$~GeV/$c^{2}$ and $m_{\Zn}=91.19$~GeV/$c^{2}$~\cite{pdg}.
They are mainly produced in the hard partonic scattering processes with large momentum transfer at the early stage of the (ultra-)relativistic nucleus--nucleus collisions.
Their main production processes are
\begin{equation*}
u\overline{d}\rightarrow\Wp, \hspace{0.5cm} d\overline{u}\rightarrow\Wm
\end{equation*}
and
\begin{equation*}
u\overline{u}\rightarrow\Zn, \hspace{0.5cm} d\overline{d}\rightarrow\Zn
\end{equation*}
in the leading order approximation~\cite{martin}.
With the involvement of the valence $u$- and $d$-quarks inside the beam particles, the isospin difference between protons and neutrons is important on the production of $\Wp$ and $\Wm$.
It is therefore expected that the charge asymmetry between $\Wp$ and $\Wm$ is related to the relative abundance of protons or neutrons of the colliding nuclei, irrespective of the collision system size.

In comparison with the evolution time of the heavy-ion collision system, $10$ to $100$~fm/$c$ for instance,
decay time of $\WZ$, which can be estimated with the full decay width $\Gamma$ \cite{pdg},
\begin{eqnarray*}
t = \frac{\hbar}{\Gamma},\hspace{0.4cm}
t_{\Wpm} = 0.0922~{\rm fm/}c,\hspace{0.4cm}
t_{\Zn} = 0.0791~{\rm fm/}c,
\end{eqnarray*}
is very short.
The $\WZ$ leptonic decays
\begin{equation*}
\Wp\rightarrow l^{+}\nu_{l}, \hspace{0.5cm} \Zn\rightarrow l^{+}l^{-},~(l{\rm :}~e,\mu,\tau)
\end{equation*}
are nearly instantaneous.
As the produced leptons weakly interact with the partonic and hadronic matter created in heavy ion collisions, $\Wpm$ and $\Zn$,
similar as the prompt direct photons, are powerful probes for investigating the properties of the
initial stage of the evolving system and the partonic structure of the colliding nuclei.

The measurement of $\WZ$ production is very important to calibrate the
collision geometry information of nuclear collisions and of great experimental
interests. Many observables in the heavy ion studies require the knowledge of
geometric quantities such as the nuclear thickness function $\avg{\TAA}$, the
number of participant nucleons $\avg{\Npart}$, and the number of binary
collisions $\avg{\Ncoll}$, which are usually obtained using the Glauber model
\cite{shor,abel,abel1,misk}.
Without participating the final state interactions, $\WZ$ bosons are ideal probes to test the validity of collision geometries applied in the experimental analysis.

The CMS and ATLAS Collaborations have first measured $\Wpm$ and $\Zn$
production in Pb--Pb collisions at $\sNN=2.76$~TeV
\cite{cms1,atlas1,cms2,atlas2}. Recently, the ALICE and ATLAS Collaborations
published the measurements of $\Zn$ production at forward rapidities
\cite{alice1} and $\WZ$ production at mid-rapidity~\cite{atlas3,atlas4}, in
Pb--Pb collisions at $\sNN=5.02$~TeV, respectively. The $\Wpm$ and $\Zn$
production cross sections are also measured in p--Pb collisions at $\sNN=5.02$
by the ALICE and CMS collaborations \cite{alice2,cms3}.
All those measurements are declared to be well reproduced by the leading-order
(LO) and/or next-to-leading-order (NLO) perturbative Quantum Chromo Dynamics
(pQCD) calculations~\cite{ct14,eps09,cteq15,epps} using the CT14 Parton
Distribution Function (PDF) set \cite{ct14} with and without the parameterized
nuclear modified PDF (nPDF) like EPPS16~\cite{epps}. As the experimental data
analysis relies strongly on the template calculated with NLO pQCD
\cite{powheg}, comparing experimental data to the LO or NLO pQCD predictions
only is incomplete. The study of $\WZ$ production in heavy-ion collisions with
Monte-Carlo simulation may provide more differential understandings into the
microscopic transport properties of the collision system.

\section {Model}
A parton and hadron cascade model PACIAE~\cite{sa1} is employed to simulate $\Zn$ production in pp and Pb--Pb collisions at centre-of-mass energy $5.02$~TeV.
The results are compared with the ALICE measurements~\cite{alice1}.
The $\Wpm$ production is predicted in Pb--Pb collisions at $\sNN=5.02$~TeV.
A systematic analysis of $\Wpm$ charge asymmetry in nucleon--nucleon (NN): pp,
pn, np, nn collisions and in nucleus--nucleus (AA): Cu--Cu, Au--Au, Pb--Pb
and U--U collisions is also studied.

\begin{widetext}
\begin{table}
\centering
\begin{varwidth}{\textwidth}
\caption{Per-event multiplicity of $\Zn$ and $\Wpm$ bosons in rapidity interval $2.5<y<4.0$ for three different centrality classes (\cent{0}(20), \cent{20}{90} and \cent{0}{90}) in Pb--Pb collision at $\sNN=5.02$~TeV.}
\end{varwidth}

\begin{tabularx}{\textwidth}{@{} CCCC @{}}
\hline
\hline \\
Centrality & \cent{0}{20} & \cent{0}{90} & \cent{20}{90} \\
\hline \\
$\dd N_{\Zn}/\dd y|_{exp.}$                 & $0.996\times 10^{-7}$ & $0.379\times 10^{-7}$ & $0.198\times 10^{-7}$ \\
\hline \\
Rescaled $\dd N_{\Zn}/\dd y|_{exp.}^{**}$   & $5.03$ & $1.91$ & $1.00$ \\
\hline \\
$\dd N_{\Zn}/\dd y|_{paciae}$               & $9.92$ & $4.42$ & $2.84$ \\
\hline \\
Rescaled $\dd N_{\Zn}/\dd y|_{paciae}^{**}$ & $3.50$ & $1.56$ & $1.00$ \\
\hline \\
$\dd N_{\Wp}/\dd y|_{paciae}$               & $4.97$ & $2.23$ & $1.43$ \\
\hline \\
Rescaled $\dd N_{\Wp}/\dd y|_{paciae}^{**}$ & $3.49$ & $1.56$ & $1.00$ \\
\hline \\
$\dd N_{\Wm}/\dd y|_{paciae}$               & $5.30$ & $2.37$ & $1.50$ \\
\hline \\
Rescaled $\dd N_{\Wm}/\dd y|_{paciae}^{**}$ & $3.54$ & $1.58$ & $1.00$ \\
\hline
\hline \\
\multicolumn{4}{l}{* Taken from ALICE~\cite{alice1}} \\
\multicolumn{4}{l}{** Rescale to the digital in 20-90\% centrality.}
\end{tabularx}
\label{tab:systpPb}
\end{table}
\end{widetext}

\begin{widetext}
\begin{center}
\begin{figure}[htbp]
\centering
\hspace{-0.50cm}
\includegraphics[width=0.3\textwidth]{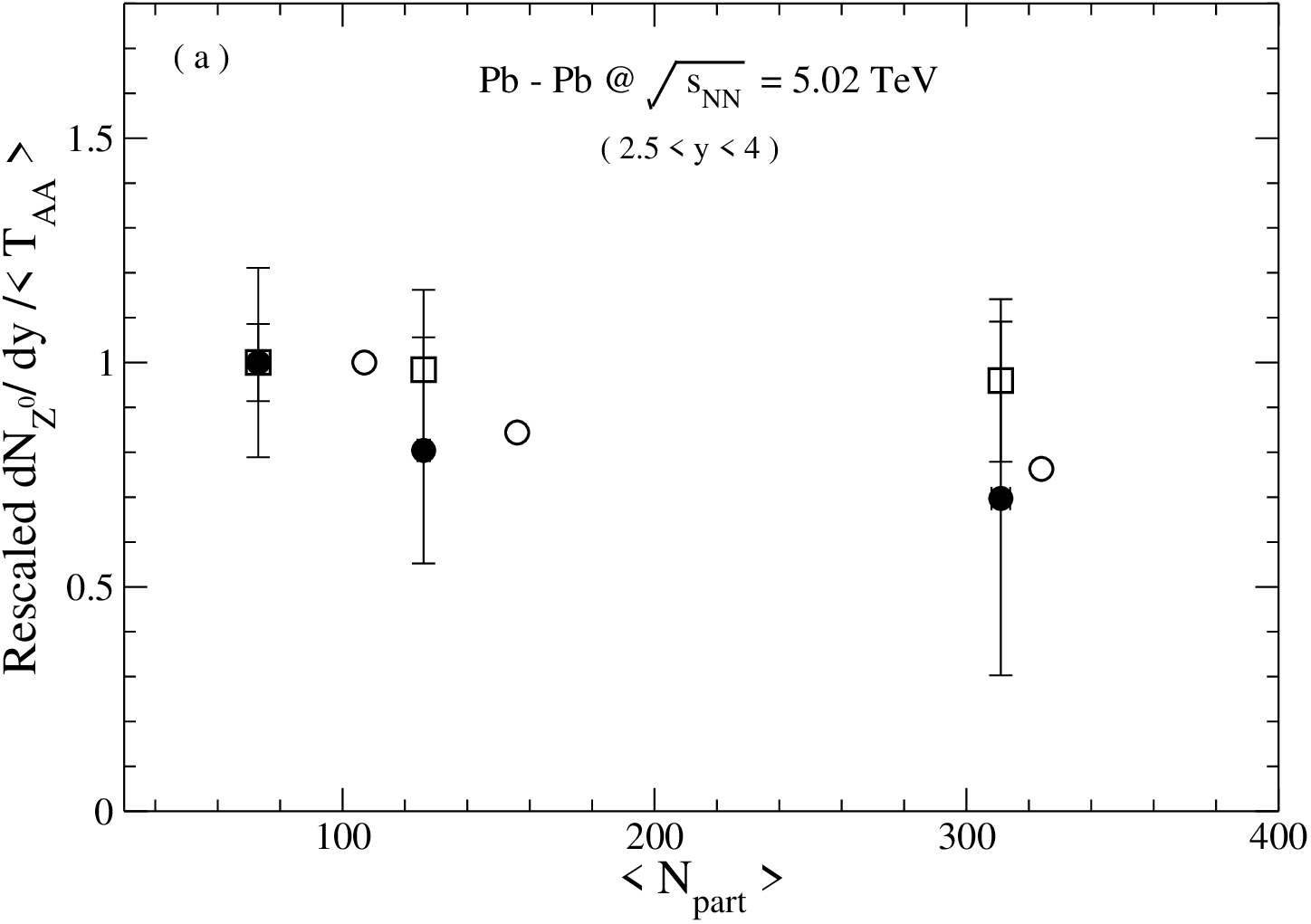} \hspace{0.1cm}
\includegraphics[width=0.3\textwidth]{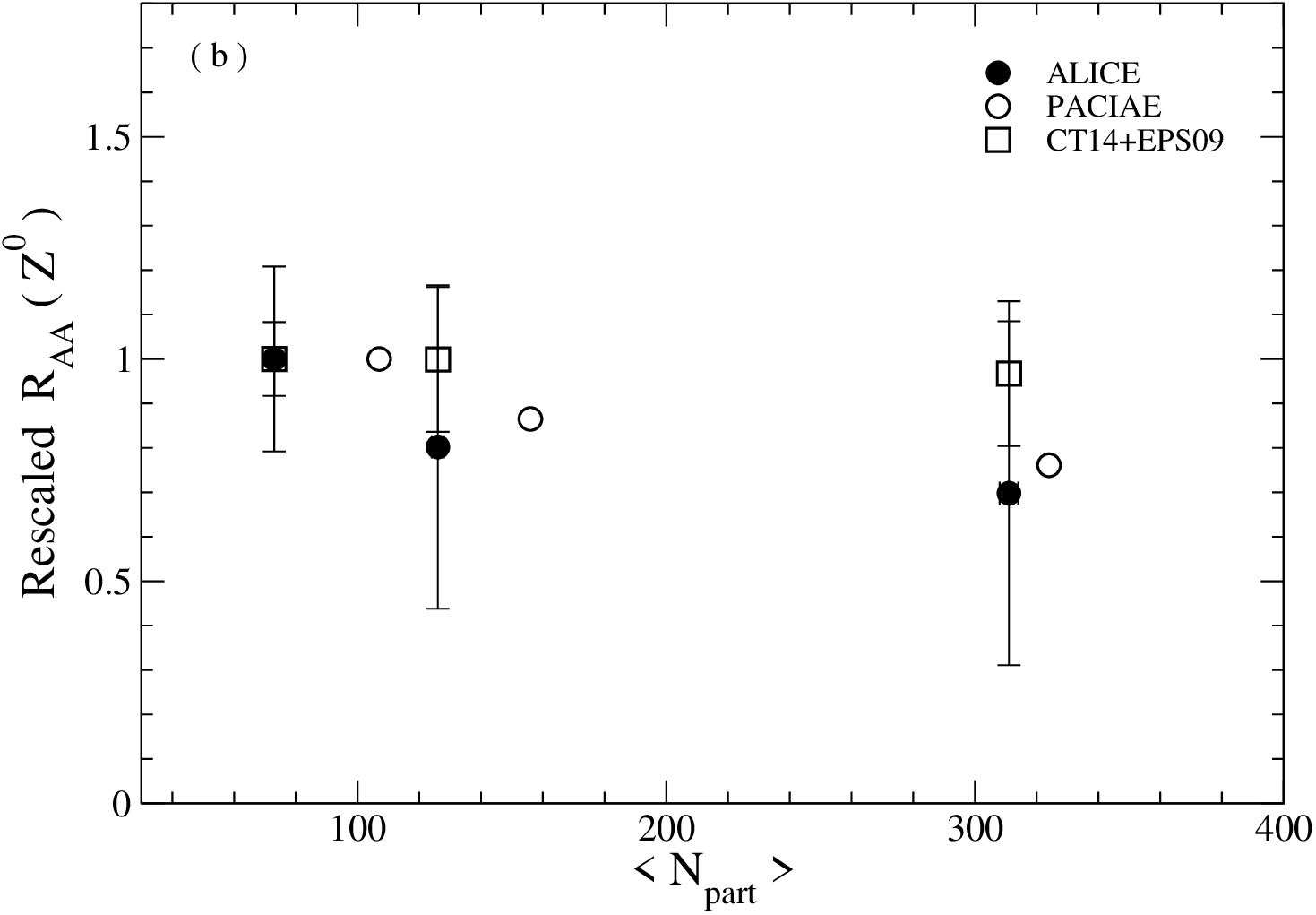} \hspace{0.1cm}
\includegraphics[width=0.3\textwidth]{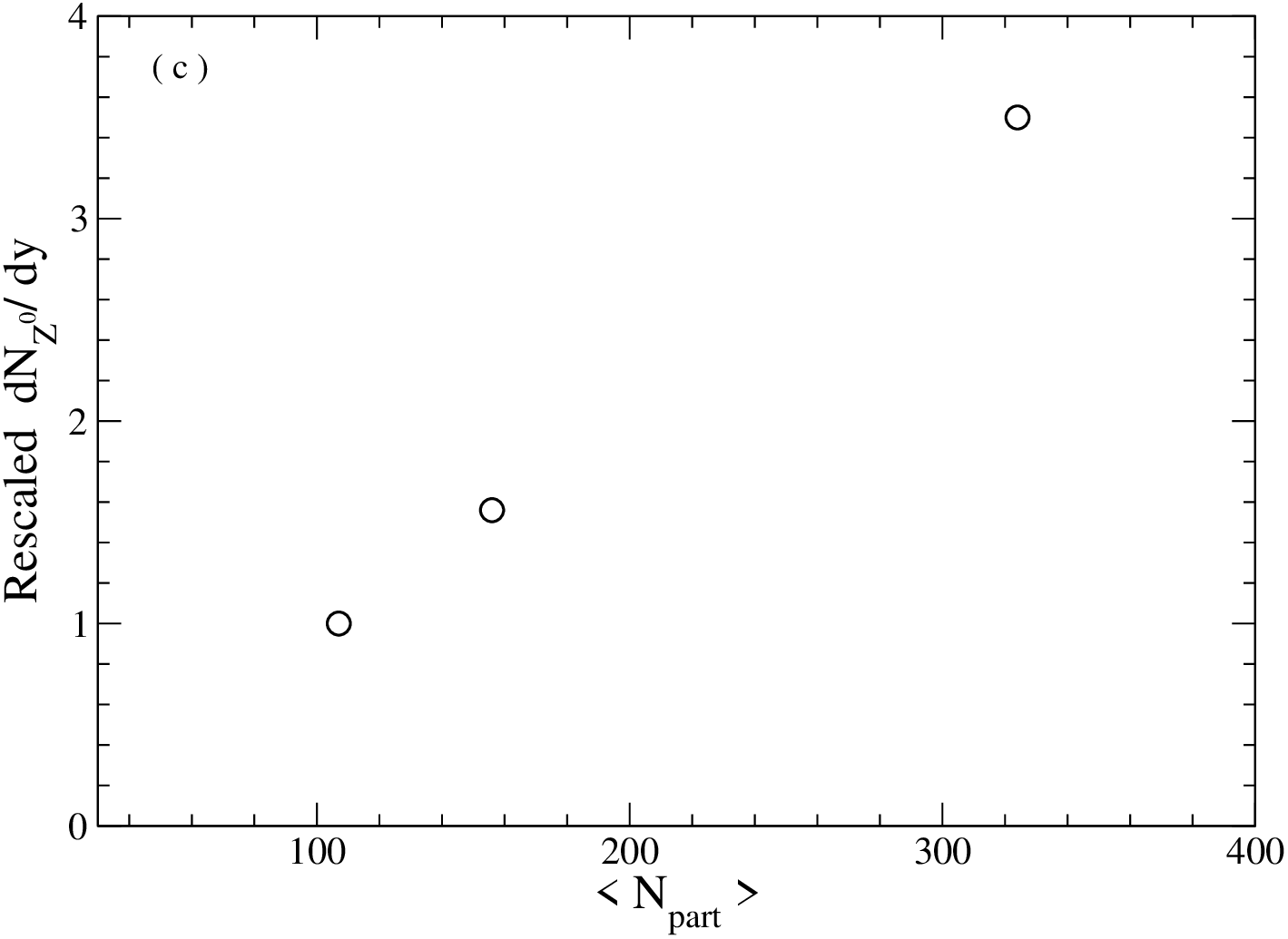}
	\caption{Centrality dependance on $\Zn$ production in
	Pb--Pb collisions at $\sNN=5.02$~TeV: panel (a) for
	$\dndy$~/~$\avg{\TAA}$, (b) for $\RAA$, and (c) for $\dndy$.}
\label{zc}
\end{figure}
\end{center}
\end{widetext}

The PACIAE model is based on \PYTHIA\ event generator (version $64.28$)
\cite{soj1}. For NN collisions, with respect to \PYTHIA, the partonic and
hadronic rescatterings are introduced before string formation and
after the hadronization, respectively. The final hadronic state is developed
from the initial partonic hard scattering and parton showers, followed by
parton rescattering, string fragmentation, and hadron rescattering stages.
Thus, the PACIAE model provides a multi-stage transport description on the
evolution of the collision system.
\begin{widetext}
\begin{center}
\begin{figure}[htbp]
\centering
\hspace{-0.50cm}
\includegraphics[width=0.3\textwidth]{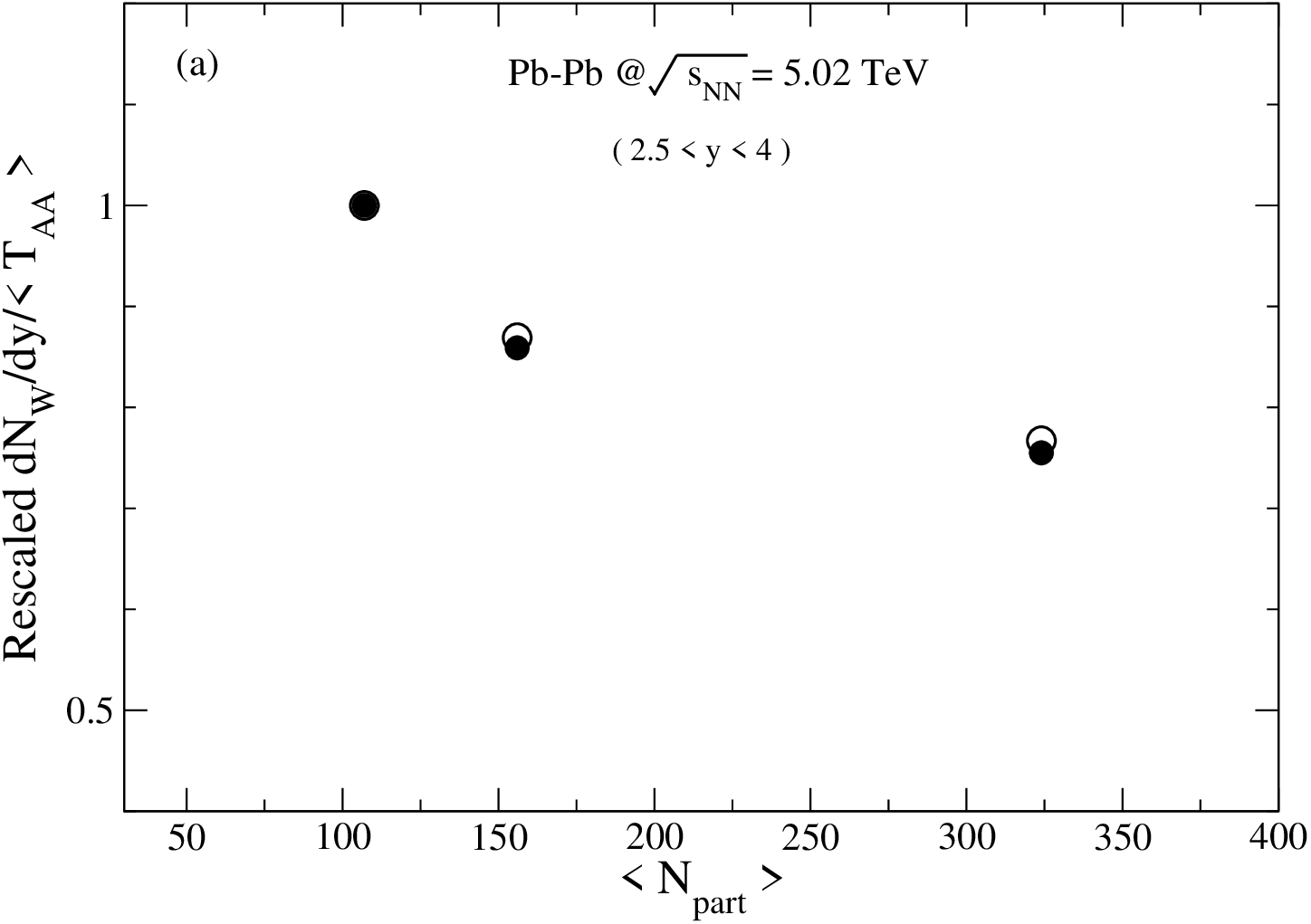} \hspace{0.1cm}
\includegraphics[width=0.3\textwidth]{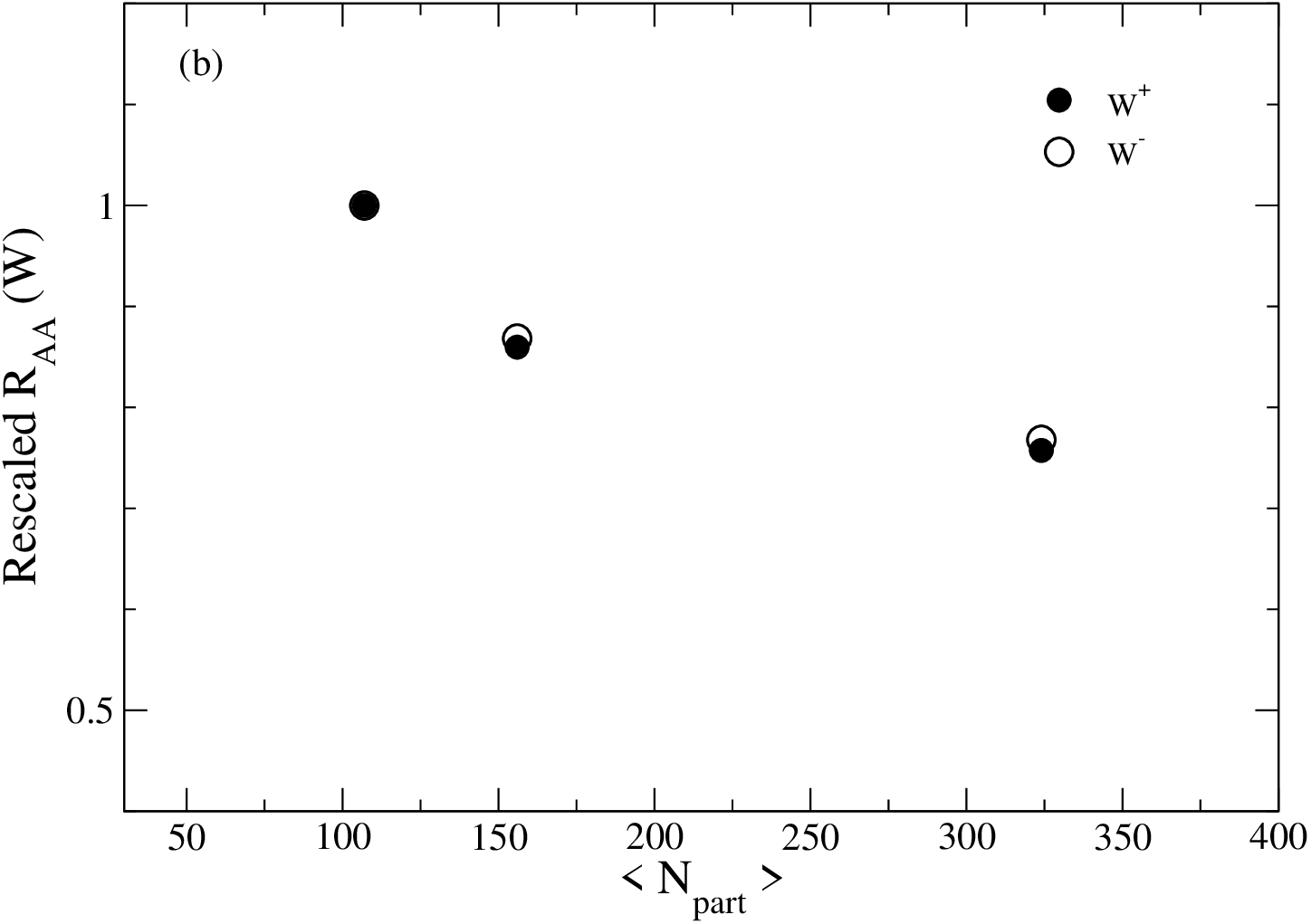} \hspace{0.1cm}
\includegraphics[width=0.3\textwidth]{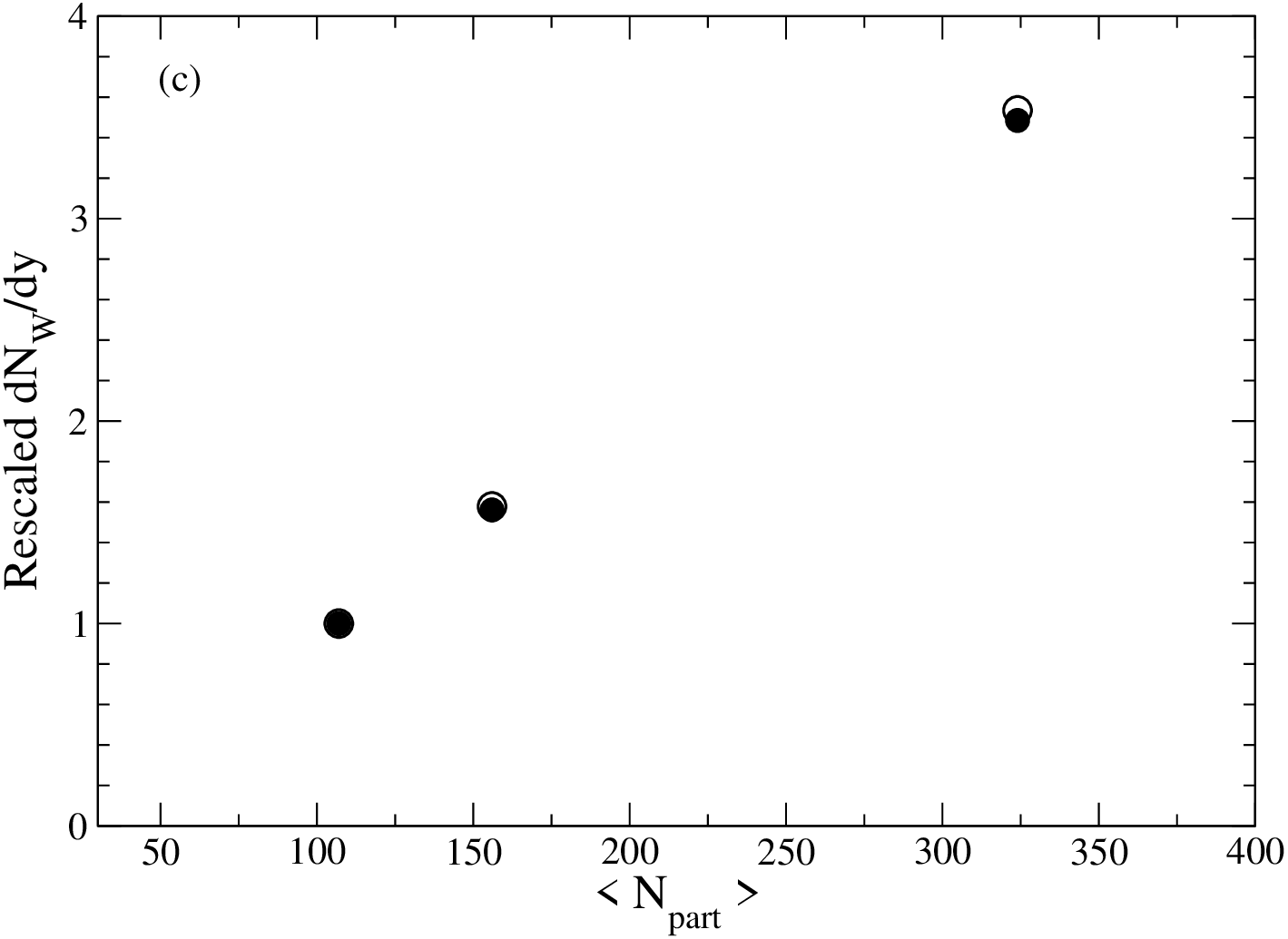}
\caption{Centrality dependance on $\Wpm$ productions in Pb--Pb collisions at
	$\sNN=5.02$~TeV: panel (a) for $\dndy$~/~$\avg{\TAA}$, (b) for
	$\RAA$, and (c) for $\dndy$.}
\label{wc}
\end{figure}
\end{center}
\end{widetext}
For AA collisions, the initial positions of nucleons in the colliding nucleus
are sampled according to the Woods-Saxon distribution. Together with the
initial momentum setup of $p_{x} = p_{y} = 0$ and $p_{z} =p_{\rm beam}$ for
each nucleon, a list containing the initial state of all nucleons in a given
AA collision is constructed. A collision happened between two nucleons
from different nuclei if their relative transverse distance is less than or
equal to the minimum approaching distance:
$D\leq\sqrt{\sigma_{\rm NN}^{\rm tot}/\pi}$.
The collision time is calculated with the assumption of straight-line trajectories. All such nucleon pairs compose a NN collision time list.
A NN collision with least collision time is selected from the list and executed by \PYTHIA\ (PYEVNW subroutine) with the hadronization temporarily turned-off, as well as the strings and diquarks broken-up. The nucleon list and NN collision time list are then updated.
A new NN collision with least collision time is selected from the updated NN collision time list and executed with repeating the aforementioned steps till the
NN collision list is empty.

With those procedures, the initial partonic state for a AA collision is constructed.
Then, the partonic rescatterings are performed, where the LO-pQCD parton-parton cross section~\cite{ranft,field} is employed.
After partonic rescatterings, the string is recovered and then hadronized with the Lund string fragmentation scheme resulting in an intermediate hadronic state.
Finally, the system proceeds into the hadronic rescattering stage and produces the final hadronic state observed in the experiments.

The per-event $\WZ$ production yield is very low, e.g., $\dd N_{\Zn}/\dd y\sim 10^{-7}$ at forward-rapidity in the most $20\%$ central Pb--Pb collisions at $\sNN= 5.02$~TeV, as shown in Tab. 1.
In this study, the relevant production channels are activated in a user controlled approach by setting ${\rm MSEL}=0$ in $\PYTHIA$ together with the following subprocesses switched on:
\begin{eqnarray*}
f_{i}\overline{f}_{j} & \rightarrow & \Wp/\Wm   \\
f_{i}\overline{f}_{j} & \rightarrow & g\Wp/\Wm  \\
f_{i}\overline{f}_{j} & \rightarrow & \gamma \Wp/\Wm \\
f_{i}g                & \rightarrow & f_{k}\Wp/\Wm   \\
f_{i}\overline{f}_{j} & \rightarrow & \Zn\Wp/\Wm     \\
f_{i}\overline{f}_{i} & \rightarrow & \Wp\Wm
\end{eqnarray*}
for $\Wpm$ production, and
\begin{eqnarray*}
f_{i}\overline{f}_{i} & \rightarrow & \gamma^{*}/\Zn    \\
f_{i}\overline{f}_{i} & \rightarrow & g(\gamma^{*}/\Zn) \\
f_{i}\overline{f}_{i} & \rightarrow & \gamma(\gamma^{*}/\Zn) \\
f_{i}g                & \rightarrow & f_i(\gamma^{*}/\Zn)    \\
f_{i}\overline{f}_{i} & \rightarrow & (\gamma^{*}/\Zn)(\gamma^{*}/\Zn) \\
f_{i}\overline{f}_{j} & \rightarrow & \Zn\Wp/\Wm
\end{eqnarray*}
for $\Zn$ production. In aforementioned equations $f$ refers to fermions
(quarks) and its subscript stands for flavor code.

The production of the $\WZ$ boson is performed using a triggered simulation approach. A normalization factor
is needed to account for the trigger bias effect.
Unfortunately, this normalization factor is not able obtaining from default
\PYTHIA\ run where the $\WZ$ production subprocesses were not active. The
comparison between results using bias sampling technique and experimental data is available investigating trend of the observable only.
Therefore we apply a self-normalized rescaling technique to both of the
experimental observables and the model calculations with respect to a
reference value at a given argument (such as $\Npart$ in Fig.~\ref{zc}), as
indicated in Tab. 1 and Fig.~\ref{zc}.

As the Quark Gluon Matter and
Hadronic Matter are both nearly transparent to the $\WZ$ bosons, the
rescattering of $\WZ$ bosons are not considered in the partonic and hardonic
evolutions in the PACIAE simulations. Thus the results of $\WZ$ productions
from PACIAE simulations are nearly the same as the ones in pQCD for pp
collisions.

\begin{figure}[htbp]
\includegraphics[width=0.3\textwidth]{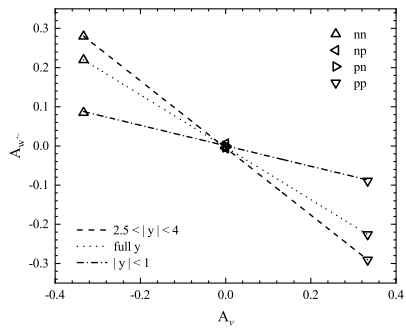}
\caption{The correlation between boson ($\Wpm$) and fermion (valence quark) charge asymmetry in the elementary nuclear-nuclear
collisions: n--n (triangle-up); n--p (triangle-left); p--n (triangle-right); p-p (triangle-up)
at $\sqrt{s_{NN}}$=5.02 TeV in three symmetrical y intervals: $|y|<1$, $2.5<|y|<4.0$ and full $y$ phase space.
The linear fitting are shown for those $y$ intervals as well.}
\label{Aw-np}
\end{figure}

\begin{figure}[htbp]
\includegraphics[width=0.3\textwidth]{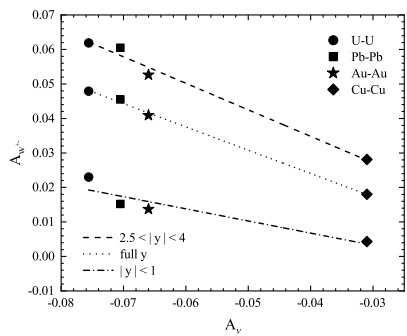}
\caption{The correlation between boson ($\Wpm$) and fermion (valence quark) charge asymmetry in symmetrical nucleus-nucleus collisions: U--U (full circle), Pb--Pb (full square), Au--Au (star), Cu--Cu (full diamond) at $\sqrt{s_{NN}}$ = 5.02 TeV in three symmetrical y intervals: $|y|<1$, $2.5<|y|<4.0$ and full $y$ phase space. The linear fitting are shown for those $y$ intervals as well.}
\label{Aw-AB}
\end{figure}

\section {Results and Discussions}
The comparison of the rescaled $\Zn$ rapidity-differential density
$\dd N_{\Zn}/\dd y~/\avg{\TAA}$ as a function of ${\Npart}$ between PACIAE
simulations and the ALICE measurements is shown on the Fig.~\ref{zc} panel (a)
in Pb--Pb collisions at $\sNN=5.02$~TeV. In this figure the \cent{0}{20}
centrality class is the reference point. The panel (a) shows that ALICE
measurements~\cite{alice1} are well reproduced by the PACIAE simulations within
uncertainties and the centrality dependence is not strong. Similarly,
Fig.~\ref{zc} panel (b) gives the centrality dependence of $\RAA(\Zn)$
\begin{equation}
\RAA=
\frac{1}{\avg{\Ncoll}}
\frac{\dndy|_{\rm PbPb}}{\dndy|_{\rm pp}}.
\end{equation}
Again, PACIAE simulations agree with the ALICE measurements within errors.
It is interesting that the centrality dependence here is nearly the same as
the one in panel (a). Finally, Fig.~\ref{zc} panel (c) is the centrality
dependence of $\dd N_{\Zn}/\dd y$. Here the centrality dependence is stronger
than the one in panel (a) or (b), as it should be.

\begin{figure}[htbp]
\includegraphics[width=0.3\textwidth]{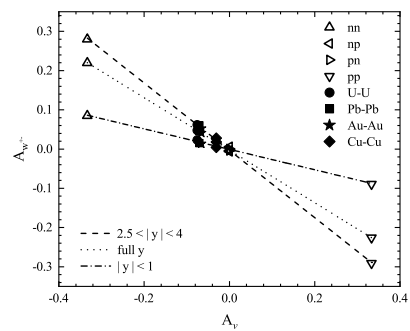}
\caption{The correlation between boson ($\Wpm$) and fermion (valence quark)charge asymmetry in the elementary nuclear-nuclear collisions and symmetrical nucleus-nucleus collisions.}
\label{Aw-tot}
\end{figure}

A similar model prediction for $\Wpm$ production in Pb--Pb collisions at
$\sNN=5.02$~TeV is shown in Fig.~\ref{wc}. It is found that the centrality dependence shown in Fig.~\ref{wc} is
quite close to the one in Fig.~\ref{zc}.

The asymmetry between $\Wp$ and $\Wm$ productions stemming from
the isospin effect can be studied with the $\Wpm$ charge asymmetry
\begin{equation}
A_{\rm \Wpm}=\frac{N_{\Wm}-N_{\Wp}}{N_{\Wm}+N_{\Wp}}.
\end{equation}
as a function of colliding system valence quark charge asymmetry
\begin{equation}
A_v=\frac{N_u-N_d}{N_u+N_d},
\end{equation}
where $N_u$ ($N_d$) stands for the total number of valence $u$ ($d$) quark in the colliding system.

The correlation between $\Wpm$ and valence quark charge asymmetry in the
elementary nucleon--nucleon collisions: n--n (triangle-up), n--p (triangle-left),
p--n (triangle-right) and p--p (triangle-down) at $\s=5.02$~TeV at different $y$
intervals: $|y|<1$, $2.5<|y|<4.0$, and full y phase space are shown in
Fig.~\ref{Aw-np}. All those results in three $y$ intervals can be fitted by
a linear function with different slopes. We can also find a feature that the
positive correlation flips into negative correlation from n--n to n--p to p--n and
to p--p in these three different rapidity intervals. It shows that the slope
obtained at mid-rapidity is gentler than that at forward rapidities.
This is due to that the isospin effect on $\Wpm$ production is more pronounced at forward rapidities~\cite{zconesa}.

In Fig.~\ref{Aw-AB} the correlation between $\Wpm$ and valence quark charge
asymmetries in minimum bias symmetrical nucleus-nucleus: U--U (full circle),
Pb--Pb (full square), Au--Au (star), Cu--Cu (full diamond) collisions at
$\s=5.02$~TeV in three different $y$ intervals:$|y|<1$, $2.5<|y|<4.0$, and
full y phase space are given. The linear fitted results are shown as well.
Also different $y$ interval has different fitted slope.
The valence quark charge asymmetry factor $A_v$ are all negative, while the $\Wpm$ charge asymmetry factor $A_{\rm \Wpm}$
are all positive, that is to say the correlation have no sign-change form U--U to Pb--Pb to Au--Au and to Cu--Cu
collision systems. If we put the data from Fig.~\ref{Aw-np} and
Fig.~\ref{Aw-AB} together, as shown in Fig.~\ref{Aw-tot}, we can get very good
linear correlations for all collision systems at different y intervals.

\section {Summary}
The parton and hadron cascade model PACIAE is employed to simulate the dynamical production of $\Wpm$ and $\Zn$ bosons in nucleon--nucleon (NN) and nucleus--nucleus (AA) collisions for the first time.
The rescaled $\dndy/\avg{\TAA}$ and $\RAA$ for $\Zn$ bosons measured by ALICE in p--p and Pb--Pb collisions at centre-of-mass energy $5.02$~TeV~\cite{alice1} are reproduced.
The correlations between $\Wpm$ charge asymmetry ($A_{\Wpm}$) and valence $u$- and $d$-quark number asymmetry ($A_{\nu}$) are studied in three rapidity intervals (mid-rapidity $\abs{y}<1$, forward rapidities $2.5 < \abs{y} < 4$, and the full rapidity region).
In each rapidity interval, a linear scaling behavior between $A_{\Wpm}$ and $A_{\nu}$ is observed by combing the results in different elementary NN (p--p, p--n, n--p and n--n) collision systems.
Results obtained from various AA (Cu--Cu, Au--Au, Pb--Pb and U--U) collision systems well follow the scaling established by NN collisions.
The nuclear-modified PDF (nPDF) is not adopted in the AA collision setup in this analysis.
And the dynamic shadowing on nucleon implemented in PACIAE simulations may
affect only the probability of binary collisions in a given AA collision, but
not the kinematics of the hard-scattering out-going particles (e.g., $\WZ$).
The universal scaling behavior between $A_{\Wpm}$ and $A_{\nu}$ observed in
both NN and AA collision systems hence may provide a vacuum (without the
nuclear modification) baseline of the isospin effect on $\Wpm$ production. The
nuclear modification on the vacuum nucleon PDF may change the in-going parton
kinematics for the hard processes and consequently may impact the $\pT$- and
$y$-distribution of the out-going particles. This effect may result in a
breaking of the $A_{\Wpm}$--$A_{\nu}$ scaling established in vacuum if looking
into a given $\Wpm$ $\pT$--$y$ phase space. Therefore, by measuring $\pT$ and
$y$ differentially, the deviation of the $A_{\Wpm}$--$A_{\nu}$ correlations
between NN and AA systems should provide a new tool to constrain the nPDF in a
wide Bjorken-$x$ region. This study serves as a benchmark for the research on
the initial conditions of heavy-ion collisions and also sheds light on the
understanding of medium-induced higher-order effects on $\WZ$ production in
future works.

\begin{acknowledgments}
This work was supported by the National Natural Science Foundation of China
(11775094, 11805079, 11905188, 11775313), the Continuous Basic Scientific
Research Project (No.WDJC-2019-16) in CIAE, National Key Research and
Development Project (2018YFE0104800) and by the 111 project of the foreign
expert bureau of China.
\end{acknowledgments}

\end{document}